\documentclass[aps,prb,reprint]{revtex4-2}

\usepackage{graphicx}
\usepackage{dcolumn}
\usepackage{bm}
\usepackage{siunitx}
\usepackage{color}
\usepackage{soul}

\begin{document}

\title{Structural dynamics of polycrystalline graphene}
\author{Zihua Liu}
\email{z.liu1@uu.nl}
\altaffiliation[Also at ]{Department of Information and Computing Sciences,
	Utrecht University, Princetonplein 5,
	3584 CC Utrecht, The Netherlands}
\author{Debabrata Panja}
\affiliation{Department of Information and Computing Sciences,
	Utrecht University, Princetonplein 5,
	3584 CC Utrecht, The Netherlands}
\author{Gerard T. Barkema}
\affiliation{Department of Information and Computing Sciences,
	Utrecht University, Princetonplein 5,
	3584 CC Utrecht, The Netherlands}

\date{\today}

\begin{abstract}
  The exceptional properties of the two-dimensional material graphene
  make it attractive for multiple functional applications, whose
  large-area samples are typically polycrystalline. Here, we study the
  mechanical properties of graphene in computer simulations and
  connect these to the experimentally relevant mechanical
  properties. In particular, we study the fluctuations in the lateral
  dimensions of the periodic simulation cell. We show that over short
  time scales, both the area $A$ and the aspect ratio $B$ of the
  rectangular periodic box show diffusive behavior under zero external
  field during dynamical evolution, with diffusion coefficients $D_A$
  and $D_B$ that are related to each other. At longer times,
  fluctuations in $A$ are bounded, while those in $B$ are not. This
  makes the direct determination of $D_B$ much more accurate, from
  which $D_A$ can then be derived indirectly. We then show that the
  dynamic behavior of polycrystalline graphene under external forces
  can also be derived from $D_A$ and $D_B$ via the Nernst-Einstein
  relation. Additionally, we study how the diffusion coefficients
  depend on structural properties of the polycrystalline graphene, in
  particular, the density of defects.
\end{abstract}

\maketitle

\section{Introduction}

Graphite is a material in which layers of carbon atoms are stacked
relatively loosely on top of each other.  Each layer consists of
carbon atoms, arranged in a honeycomb lattice.  A single such layer is
called graphene.  This material has many exotic properties, both
mechanical and electronic.  Experimentally produced samples of
graphene are usually polycrystalline, containing many intrinsic
\cite{Yazyev2014,Rasool2014,Tison2014}, as well as extrinsic
\cite{Araujo2012} lattice defects.  Unsaturated carbon bonds are
energetically very costly
\cite{Trevethan2014,Cui2020,Ganz2017,Budarapu2015,He2014}, and
therefore extremely rare in the bulk of the material. Polycrystalline
graphene samples are therefore almost exclusively three-fold
coordinated, and well described by a continuous random network (CRN)
model \cite{Zachariasen1932}, introduced by Zachariasen almost 90
years ago.

Polycrystalline graphene is continuously evolving in time, from one
CRN-like state to another. A mechanism by which such a topological
change can happen, was introduced by Wooten, Winer, and Weaire (WWW)
in the context of the the simulation of samples of amorphous Si and
Ge. This so-called WWW algorithm became the standard modeling approach
for the dynamics of these kind of models \cite{Wooten1985,Wooten1987}.

In the WWW approach, a configuration $C_i$ consists of a list of the
coordinates of all $N$ atoms, coupled with an explicit list of the
bonds between them. From this configuration $C_i$, a trial
configuration $C'_i$ is produced via a {\it bond transposition}: a
sequence of carbon atoms $\{i,j,k,l\}$ is selected, connected with
explicit bonds $i$-$j$, $j$-$k$ and $k$-$l$. The first and last of
these bonds are then replaced by bonds $i$-$k$ and $j$-$l$, while bond
$j$-$k$ is preserved.  After this change in topology, the atoms are
allowed to relax their positions. This simulation approach requires
a potential that uses the explicit list of bonds, for instance the
Keating potential \cite{Keating1966} for amorphous silicon. The
resulting configuration is then called the trial configuration
$C'_i$. The proposed change to this trial configuration is either
accepted, i.e. $C_{i+1}=C'_i$, or rejected, i.e.  $C_{i+1}=C_i$. The
acceptance probability is determined by the energy difference via the
Metropolis criterion:
\begin{equation}
	P = \min \{ 1,\exp ( - \beta \Delta E)\},
	\label{eq1}
\end{equation}
where $\beta=(k_BT)^{-1}$, with Boltzmann constant $k_B$ and
temperature $T$, and $\Delta E=E(C')-E(C)$ is the change in energy due
to the bond transposition. In this way, the simulation produces a
Markov chain $C_0 \dots C_M$, satisfying detailed balance.

The properties of polycrystalline graphene sheets have been a topic of
intense research already for some time
\cite{Cummings2019,Zeng2020,Chen2020,Estrada2019,Park2017}. More
recently, Ma \emph{et al.}  reported that the thermal conductivity of
polycrystalline graphene films dramatically decreases with decreasing
grain size \cite{Ma2017}. The work of Gao \emph{et al.} shows that the
existence of single-vacancy point defect can reduce the thermal
conductivities of graphene \cite{Gao2017}. Wu \emph{et al.} reported
the magnetotransport properties of zigzag-edged graphene nanoribbons
on an \emph{h}-BN substrate \cite{Wu2018}. Additionally, strain
effects on the transport properties of triangular and hexagonal
graphene flakes were studied in the work of Torres \emph{et al.}
\cite{Torres2018}.

This article reports on the dynamical properties of polycrystalline
graphene.  In particular, we study two geometric quantities that are
readily accessible in computer simulations without having a clear
experimental counterpart. In our simulations, the $L_x\times L_y$
graphene sample is rectangular, with periodic boundary conditions in
the $x$- and $y$-directions; the quantities of interest are the area
$A=L_xL_y$ and the aspect ratio $B=L_x/L_y$, and their mean
square displacements (MSDs) under simulations in which the dynamics is
the WWW algorithm. The results show that in the absence of external
forces, MSD$_A$ and MSD$_B$ initially both increase linearly in time.
At longer times, MSD$_A$ saturates due to geometric limitations, while
MSD$_B$ keeps increasing linearly at all times. We measure the
diffusion coefficients $D_A$ and $D_B$, and demonstrate that the two
are related.

We then continue to show that $D_A$ and $D_B$ govern the response of
the sample to stretching and shear forces respectively, following the
Nernst-Einstein relation.

The main relevance of the research presented here lies in establishing
the relation between observables that are readily accessible in
simulations but without a clear experimental counterpart ($A$ and $B$
and their dynamics), and mechanical properties of real-life graphene
(e.g.  response to external stretching and shear
forces). Additionally, we demonstrate a clear relation between MSD$_A$
and MSD$_B$, thereby also relating the bulk- and the shear-properties.
Thus far, much less is known about this shape fluctuation-driven
diffusive behavior; our work provides insight into the dynamics and
mechanics of polycrystalline graphene.

\section{The model}

For simulating graphene, we use a recently developed effective
semiempirical elastic potential \cite{Jain2015}:
\begin{eqnarray}
	E_0=\frac{3}{16}\frac{\alpha}{d^2}\sum_{i,j}(r_{ij}^2-d^2)^2
	+\frac{3}{8}\beta 
	d^2\sum_{j,i,k}\left(\theta_{jik}-\frac{2\pi}{3}\right)^2\nonumber\\
  &&\hspace{-7.2cm}+\gamma\sum_{i,jkl}r_{i,jkl}^2.
     \label{eff-potential}
\end{eqnarray}
Here, $r_{ij}$ is the distance between two bonded atoms,
$\theta_{jik}$ is the angle between the two bonds connecting atom $i$
to atoms $j$ and $k$, and $r_{i,jkl}$ is the distance between atom $i$
and the plane through the three atoms $j$, $k$ and $l$ connected to
atom $i$.  The parameter $\alpha=26.060$~eV/\AA$^{2}$ controls
bond-stretching and is fitted to the bulk modulus,
$\beta=5.511$ eV/\AA$^2$ controls bond-shearing and is fitted to the
shear modulus, $\gamma=0.517$~eV/\AA$^{2}$ describes the stability of
the graphene sheet against buckling, and $d = 1.420$ \AA~is the ideal
bond length for graphene. The parameters in the potential
(\ref{eff-potential}) are obtained by fitting to DFT calculations
\cite{Jain2015}.

This potential has been used for the study of various mechanical
properties of single-layer graphene, such as the vibrational density
of states of defected and polycrystalline graphene \cite{Jain2015a} as
well as of various types of carbon nanotubes \cite{Jain2017}, the
structure of twisted and buckled bilayer graphene \cite{Jain2016}, the
shape of nanobubbles trapped under a layer of graphene
\cite{Jain2017}, and the discontinuous evolution of defected graphene
under stretching \cite{DAmbrosio2019}.

The initial polycrystalline graphene samples are generated as in
\cite{Aurenhammer2000}.  Here, $N/2$ random points are placed in a
square simulation box with periodic boundary conditions, and the
Voronoi diagram is generated: around each random point, its Voronoi
cell is the region in which this random point is nearer than any other
random point.  We then translate the boundaries between neighboring
Voronoi cells into bonds, and the locations where three boundaries
meet into atomic positions. In this way, we have created a three-fold
coordinated CRN which is homogeneous and isotropic (i.e. does not have
preferred directions). It is, however, an energetically unfavorable
configuration; therefore, we then evolve the sample using the improved
bond-switching WWW algorithm to relax it, while preserving crystalline
density.

Up to this point, the sample is completely planar (i.e., all
$z$-coordinates are zero). After some initial relaxation, we then
assign small random numbers to the $z$-coordinates followed by energy
minimization, which results in a buckled configuration. At this point,
we also allow the box lengths $L_x$ and $L_y$ to relax. We do not
relax the box lengths already in poorly relaxed samples, because then
the sheet tends to develop all kinds of unphysical structures.

\begin{figure}[h]
  \includegraphics[width=\linewidth]{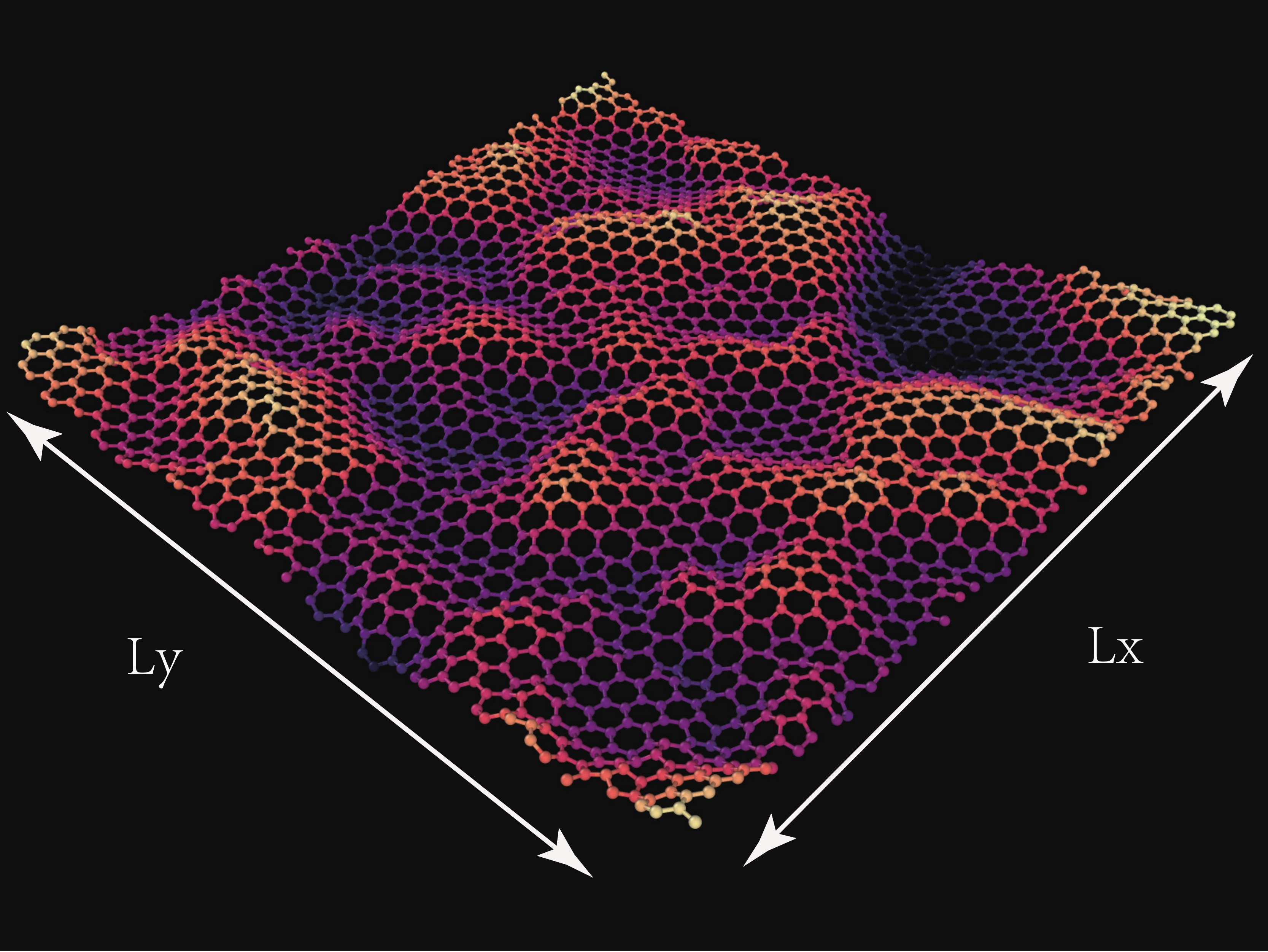}
  \caption{(color online) An initial buckled polycrystalline graphene
    sample with periodic boundary condition generated from a Voronoi
    diagram and evolved based on the WWW-algorithm. $L_x$ and $L_y$
    represent the lateral dimensions of the sample. \label{fig1}}
\end{figure}

In our implementation, we use the fast inertial relaxation engine
algorithm (FIRE) for local energy minimization \cite{Bitzek2006}; the
values of the parameters in this algorithm (${N_{\min}}$, ${f_{inc}}$,
${f_{dec}}$, ${\alpha_{start}}$ and ${f_\alpha }$) are taken as
suggested in Ref. \cite{DAmbrosio2021}. Figure~\ref{fig1} presents an
initial polycrystalline graphene sample with periodic boundary
condition generated from a Voronoi diagram and evolved based on the
WWW-algorithm.

\section{Dynamics of fluctuations in sample shapes\label{sec3}}

The oblong polycrystalline graphene sheet in our simulations has
lengths $L_x$ and $L_y$ in the $x$- and the $y$-directions
respectively, as shown in Fig. \ref{fig1}.  These are not fixed
quantities, but they fluctuate when bond transpositions are made.

\begin{figure}[t]
  \includegraphics[width=\linewidth]{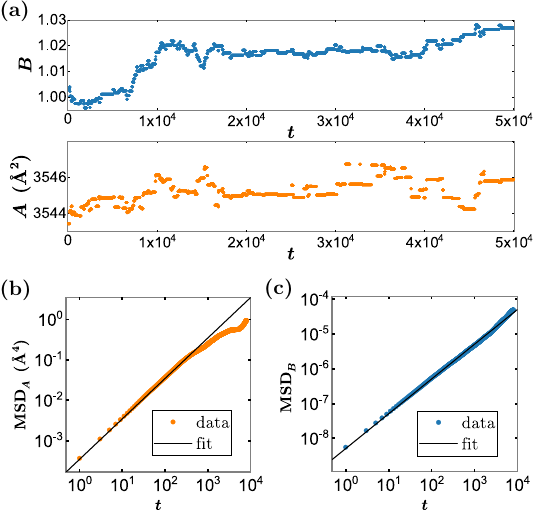}
  \caption{(color online) (a) Typical fluctuations in $A$ and $B$ in
    time for a sample with $N=1352$; note that the range of
    fluctuations in $B$ are considerably higher than in $A$. (b)
    MSD$_A(t)$ and MSD$_B(t)$ for this sample. The measured diffusion
    coefficients, as per Eq. (\ref{eq3}) are
    $D_A\approx1.737\times 10^{-4}$ \AA$^4$/[MC unit] and
    $D_B\approx2.544\times10^{-9}$ [MC unit]$^{-1}$. See text for
    details.\label{fig2}}
\end{figure}

Given that the sample is essentially two-dimensional, throughout this
paper we consider two geometric quantities defined as follows:
\begin{eqnarray}
  A(t) = L_x(t)L_y(t) \quad\mbox{and}\quad B(t) = L_x(t)/L_y(t).
  \label{eab}
\end{eqnarray}
Physically, for a flat, rectangular and homogeneous isotropic sample,
the stiffness matrix is reduced and the mechanical properties of
system can be efficiently characterized by two independent in-plane
modes due to orthorhombic symmetry, It is easiest to associate $A(t)$
and $B(t)$ to fluctuations in the sample shape in the ``bulk'' and the
``shear'' modes respectively at the macroscopic scale without these
symmetries breaking. We then track the dynamics of shape fluctuations
of the sample in terms of their mean-square displacements
MSD$_A(t)=\langle[A(t)-A(0)]^2\rangle$ and
MSD$_B(t)=\langle[B(t)-B(0)]^2\rangle$, with the angular brackets
denoting ensemble averages for a sample of fixed number of atoms and
(more or less) constant density of defects. (We will soon see that the
diffusion coefficients are functions of both these quantities.)
Characteristic fluctuations in $A$ and $B$ for a sample with 1352
atoms are shown in Fig. \ref{fig2} panel (a), and correspondingly,
their MSDs are shown in panels (b) and (c). Therein we find that
fluctuations in $A$ are relatively much smaller in magnitude than
those in $B$. Intuitively this makes sense, since relaxations through
the shear mode is energetically much more favorable than through the
bulk mode. This is also reflected in the MSDs. After a linear increase
in time, MSD$_A$ saturates at longer times, while MSD$_B$ increases
linearly at all times.  From the data for MSD$_A$ before it saturates,
and MSD$_B$ at all times, we identify the diffusion coefficients $D_A$
and $D_B$, obtained from fitting the data to the relation given by
\begin{equation}
  \mbox{MSD}(t) = 2 Dt.
  \label{eq3}
\end{equation}
Since time is measured in MC units (bond transposition moves are being
attempted once per unit of MC time), and length is measured in \AA,
the units of $D_A$ and $D_B$ are \AA$^4$/[MC unit] and [MC
unit]$^{-1}$ respectively. Time all throughout the paper is measured
in MC units.

\subsection{$D_B$ increases linearly with defect
  density\label{sec3a}}

An interesting question is what determines $D_B$ for a sample with a
given number of atoms $N$. As we expect $D_B$ to be equal to zero for
a perfect graphene sample, our first guess is that $D_B$ might depend
on the density of defects. In our computer simulations of perfectly
three-fold coordinated networks, defects are topological, in
particular rings which are not six-fold. A convenient measure of the
defect density $\rho$ is then obtained by the number of such rings per
area. Note that rings are almost exclusively 5-, 6- and 7-fold in the
well-relaxed samples as we studied. Since 5- and 7-fold rings
generally appear and disappear in pairs, one can expect that the ratio
of N5/N7 (N represents the number of 5- for 7-fold rings) is close to
unity.
\begin{figure}[t]
  \includegraphics[width=0.8\linewidth]{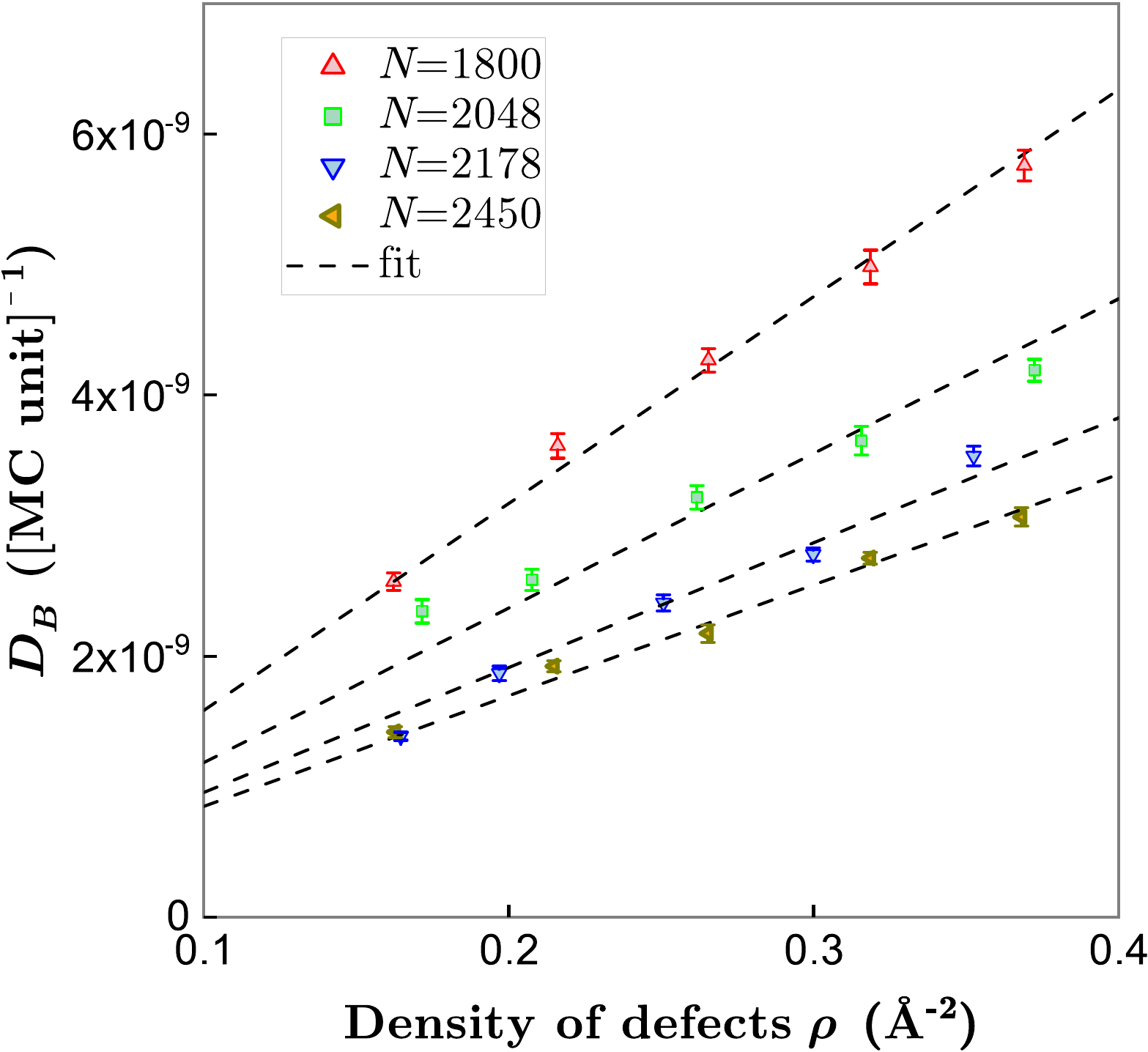}
  \caption{(color online) $D_B$ plotted for four differently-sized
    samples, each with four different defect densities (points:
    simulation data, lines: best fit passing through origin). Error
    bars represent standard error of the mean, obtained from the
    ensemble of simulation runs. See text for details. \label{fig3}}
\end{figure}

In order to test our intuition, we simulate graphene samples for four
different atom numbers (around $N=2000$), each with four different
defect densities. The results are shown in Fig.~\ref{fig3}.  Points
represent simulation data with statistical error bars, and dashes
lines are best fit lines with each line passing through the origin
(corresponding to $D_B=0$ at $\rho=0$). Even though there is no {\it a
  priori\/} reason for $D_B$ to increase linearly with $\rho$ for
every value of $N$, Fig.~\ref{fig3} demonstrates that the linear
scaling holds for the range of defect densities we simulated. Also
clear is the decreasing trend in $D_B$ with increasing $N$ for a
certain defect density.  On a technical side, each point is obtained
from averaging over 10 independent samples, and each sample is
simulated 16 times over 30,000 attempted bond transpositions at a
temperature of $kT=0.25$ eV within each run. We perform further
averaging over the initial time. The CPU time of a single attempted
bond transposition is on average 0.76 s for samples ($N$=2000).

\subsection{Relation between $D_A$ and $D_B$\label{sec3b}}

Further, since both $A$ and $B$ bear relations to $L_x$ and $L_y$, one
would expect them to be related through these length parameters, which
we establish below. In order to do so, having denoted the change in
$A$ and $B$ over a small time interval $dt$ for samples with
dimensions $L_x$ and $L_y$ by $dA$ and $dB$ respectively, we express
them in terms of small changes $dL_x$ and $dL_y$ as
\begin{eqnarray}
  \langle dA^2 \rangle &=& \left\langle \left[L_y dL_x + L_x
                           dL_y\right]^2\right\rangle
                           \quad\mbox{and}\nonumber\\
     \langle dB^2 \rangle &=&\left\langle\frac{1}{L_y^4} \left[L_y
                              dL_x - L_x dL_y\right]^2\right\rangle.        
	\label{eq4}
\end{eqnarray}
Using $\langle dL_xdL_y\rangle=0$ after an ensemble averaging,
Eq. (\ref{eq4}) leads to the simplified form
\begin{eqnarray}
  \langle dA^2\rangle&=& L_y^2\,\langle dL_x^2\rangle+L_x^2\,\langle dL_y^2\rangle\quad\mbox{and}\nonumber\\
  \langle dB^2\rangle&=&\frac{1}{L_y^4}\left[L_y^2\,\langle dL_x^2\rangle+L_x^2 \,\langle dL_y^2\rangle\right],
	\label{eq6}
\end{eqnarray}
i.e., $\langle dA^2\rangle/\langle dB^2\rangle =
L_y^4$. If we extend this analysis to finite times, for
which $L_y$ does not appreciably change, then we expect the ratio
$D_A/D_B$ to behave $\sim L^4_y$.
\begin{figure}[t]
  \includegraphics[width=0.4\textwidth]{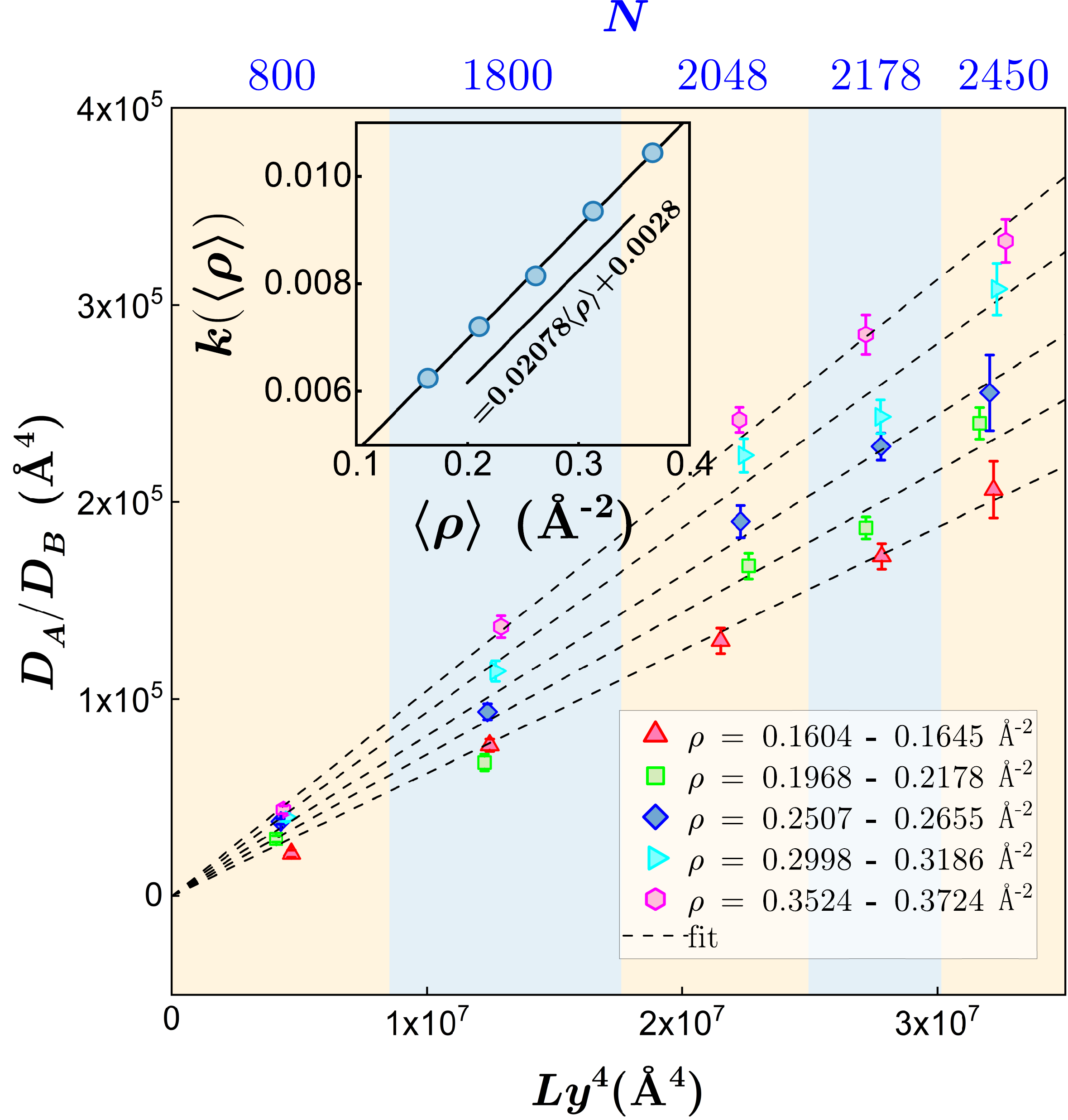}
  \caption{(color online) The ratio $D_A/D_B$ vs $L_y^4$ for different
    $N$-values and defects densities [points: simulation data, lines:
    best-fit of the form
    $D_A/D_B=k(\langle\rho\rangle)\,L^4_y$]. Error bars represent
    standard error of the mean, obtained from the ensemble of
    simulation runs. The points located a color bar are measured with
    the same $N$. The inner plot shows that $k(\langle \rho \rangle)$
    also bears a linear relation with $\langle \rho \rangle$ obtained
    from averaging in the ranges:
    $k(\langle \rho \rangle)=0.02078\langle \rho
    \rangle+0.0028$. \label{fig4}}
\end{figure}

In Fig.~\ref{fig4} we plot $D_A/D_B$ for $N$=800, 1800, 2048, 2178,
2450 and five different ranges with approximate defect densities. We
indeed observe that $D_A/D_B\sim L_y^4$: once again, simulation data are
shown as points, while the dashed lines are the best-fit
$D_A/D_B=k(\rho)\,L^4_y$ lines through the data points. The
$k$-values, summarized in Tab.~\ref{table1}, are plotted as an inset
to Fig.~\ref{fig4}. Here we determine $k$ by using statistical quantity
$\langle \rho \rangle$ obtained from averaging in the ranges, these
$k(\langle \rho \rangle)$ vs $\langle \rho \rangle$ points also lie on
a straight line, whose best-fit estimate is
$k(\langle \rho \rangle)=0.02078\langle \rho \rangle+0.0028$.
\begin{table}[!h]
  \begin{tabular}{c|c}
    $\quad\langle \rho \rangle\quad$&$k(\langle \rho \rangle)$\\
    \hline\hline
    0.16433&$\quad6.24\times10^{- 3}\quad$\\
    0.21065&$7.20\times10^{- 3}$\\
    0.26154&$8.14\times10^{- 3}$\\
    0.31323&$9.35\times10^{- 3}$\\
    0.36670&$1.04\times10^{- 2}$\\\hline
  \end{tabular}
  \caption{Values of $k$ for different values of
    $\langle \rho \rangle$, corresponding to the best-fit
    $D_A/D_B=k(\langle\rho\rangle)\,L^4_y$ lines in
    Fig.~\ref{fig4}.\label{table1}}
\end{table}

\begin{table}[!h]
	\begin{tabular}{c|c|c}
		$N$&$\langle L_{x,\text{initial}} \rangle $(\AA)& $\langle L_{y,\text{initial}} \rangle 
		$(\AA) \\
		\hline\hline
		800  & 45.35 & 45.72 \\
		1800 & 68.47 & 68.95 \\
		2048 & 73.91 & 72.58 \\
		2178 & 75.78 & 75.39 \\
		2450 & 80.15 & 80.38\\ 
		\hline
	\end{tabular}
	\caption{Dimensions of initial configurations obtained after
          optimization. Averaging for each $N$ was done over 5
          ($\langle \rho \rangle$ values listed in the table
          \ref{table1}) $\times$ 10 (independent samples) $\times$ 16
          (repetitions).\label{table2}}
\end{table}

\begin{figure}[t]
  \includegraphics[width=0.48\textwidth]{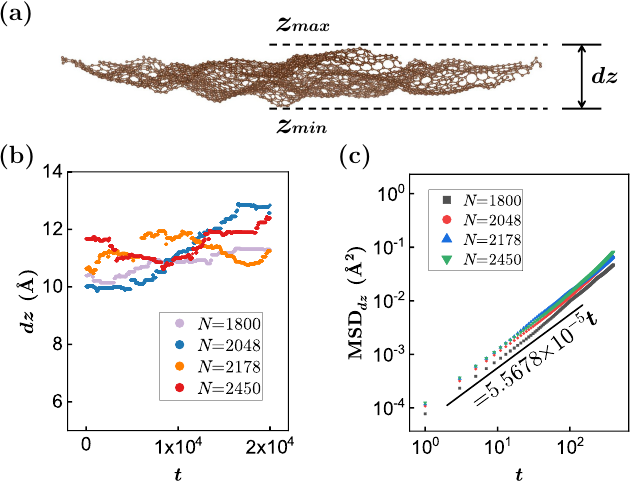}
  \caption{(color online) (a) A suspended graphene sample naturally
    tends to buckle, $dz$ is the thickness of the sample. (b) The
    variation of $dz$ for four differently-sized samples, the initial
    defects densities for all sample are fixed at around 0.15. (c)
    MSD$_{dz}$ for these samples.} \label{fig5}
\end{figure}

\subsection{MSD in the $z$-direction\label{sec3c}}

The graphene in our simulations is free-floating, and the presence of
defects causes it to buckle, i.e., the carbon atoms show displacements
in the out-of-plane direction.  During bond transpositions, the
buckling structure changes. To quantify the dynamics of buckling, we
determine the minimal and maximal values of the $z$-coordinates of the
atoms, and the difference $dz=z_{max}-z_{min}$; this is illustrated in
the top panel of Fig. 5.
\begin{equation}
  \text{MSD}_{dz}(t) = \langle[dz(t) - dz(0)]^2\rangle.
\end{equation}

Analogous to our analysis of the dynamics of $L_x(t)$ and $L_y(t)$, we
then determine the MSD$_{dz}$ of $dz(t)$.  The results for various
system sizes are shown in figure 5, in samples with a defect density
around 0.15, simulated at a temperature of $kT$ = 0.25 eV.  Figure
5(b) shows that the $dz$ fluctuates around a level $\approx11$ \AA ,
which is the typical equilibrium amplitude of the buckling for these
samples; out-of-plane displacement-related studies can be found in our
previous simulations \cite{Jain2015a}. Figure 5(c) shows that the
initial behavior is diffusive, with a diffusion coefficient that is
insensitive to $N$.

\subsection{Summary: defect density determines
  shape fluctuation dynamics \label{sec3d}}

In summary so far, we have established that the density of defects
determines $D_B$, and that the ratio $D_A/D_B=k(\langle\rho\rangle)L^4_y$ in
Sec. \ref{sec3b}. Putting these results together then implies that the
density of defects is the sole determining factor for the dynamics of
fluctuations in the sample shapes.

\section{Sample response to external forces}

That the fluctuations in quantity $B$ lead to diffusive behavior
without being limited by geometric constraints made us follow-up with
the response of the samples to externally applied forces. In
particular, if we apply a (weak) force $F_B$ to excite the shear mode,
then we expect the (linear) response in terms of ``mobility'' $\mu_B$ in the
relation $v_B=\mu_B F_B$ for the ``deformation velocity $v_B$ of the
sample along the $B$-direction'' to satisfy the Einstein relation
\begin{eqnarray}
  \mu_B = \frac{D_B}{k_BT};\quad\mbox{i.e.,}\quad v_B=\frac{D_B}{k_BT} F_B,
  \label{eq7}
\end{eqnarray}
where $k_B$ is the Boltzmann constant and $T$ is the temperature of
the sample.
\begin{figure*}
  \includegraphics[width=\linewidth]{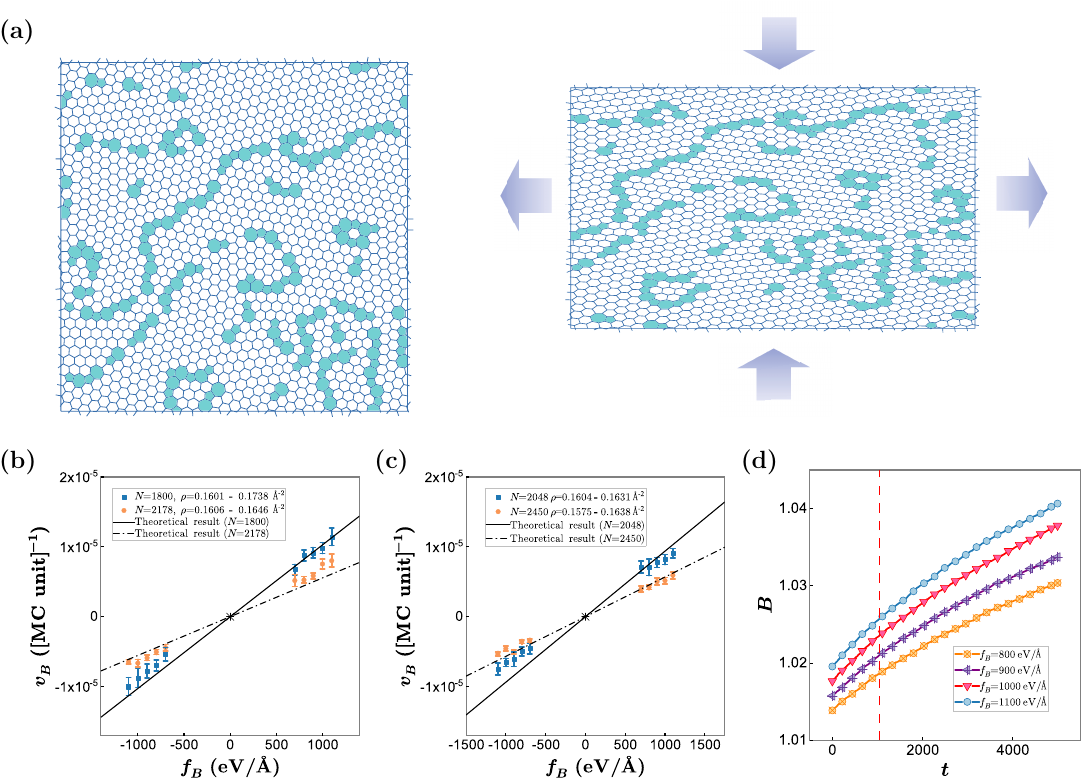}
  \caption{(color online) (a) A relaxed polycrystalline graphene
    sample. Elongated domains arise if the sample is stretched
    significantly within a short period of time. (b-c) $v_B$ directly
    measured vs. predicted by the Nernst-Einstein relation
    (\ref{eq7}). Error bars represent standard error of the mean,
    obtained from the ensemble of simulation runs. (d) Change in $B$ in
    time when a constant stretching force is applied to the sample,
    restoring tendencies of elongated domains cause slower-than-linear
    increases at longer times ($t>1000$).  \label{fig6}}
\end{figure*}

In order to check for this relation in our simulations, we add an
extra ``force term'' in the Hamiltonian in Eq. (\ref{eff-potential}),
to have the new Hamiltonian as
\begin{equation}
  E = E_0 + c\frac{L_x}{L_y} \equiv E_0 + cB,
  \label{eq8}
\end{equation}
and calculate $v_B$ in the following manner, for the applied force
$F_B=\partial E/\partial B=c$.

The behavior of the aspect ratio $B$ as a function of time, under a
constant force $f_B$, is shown in figure 6a, for forces
$f_B=\pm 800, 900, 1000$ and 1100 eV/\AA.  The curves in this figure
are obtained by averaging over 8 independent samples, each one
simulated 32 times for each value of the force.  At relatively short
times, $B$ increases linearly in time. Afterwards, the shear rate has
a tendency to slow down. We speculate that this slowing down at longer
times might be due to deformation of domains: Initially, these
crystalline domains are isotropic, but after the sample has sheared
over quite some distance, the domains become elongated. The tendency
to restore isotropy makes the sample resist further deformation.  This
is illustrated in fig. 6a. There is no a priori reason to assume that
the increase in energy due to shearing is harmonic.  In analogy to the
quartic increase of the length of a circle under this type of
deformation, we rather expect highly non-linear behavior.  At short
times, where the sample has not deformed significantly, the change in
$B$ as a response to the force $f_B$ is expected to be given by the
Nernst-Einstein equation Eq. (7).  To test this, we obtained the
short-time shear velocity $v_B$ by fitting the slopes in figure 6a for
the various forces. These measurements of $v_B$ are plotted in figures
6b and 6c, as a function of $f_B$.  Also plotted in figures 6b and 6c
are the theoretical expectations as obtained from the Nernst-Einstein
equation, in which we used the earlier obtained values for $D_B$.  The
figures 6b and 6c show agreement between the direct measurements of
$v_B$ and the theoretical expectations, indicating that with forces of
these strengths the mechanical response is well-understood.

\section{Conclusion}

Computer simulations of materials at the atomistic level usually
involve samples containing typically a few thousand atoms, with
periodic boundary conditions. Quantities that can be easily and
reliably measured in such simulations, are for instance the evolution
in time of the lateral sizes of the periodic box, such as their
fluctuations. In the simulations on graphene as presented here, the
directly observable quantities are the lateral lengths $L_x$ and $L_y$
of the rectangular periodic box. The dynamics of $L_x$ and $L_y$ are
coupled and can be better understood by considering the area
$A=L_xL_y$ and aspect ratio $B=L_x/L_y$. Specifically, we
concentrate on the mean-squared displacements of $A$ and $B$. At short
times, in which only a few atomic rearrangements occur, $A$ and $B$
show ordinary diffusive behavior, with diffusion coefficients $D_A$
and $D_B$. We show that if the changes in $L_x$ and $L_y$ are
uncorrelated, $D_A$ and $D_B$ can be obtained from each other.  While
this might not seem very surprising at first sight, it does connect
the dynamics of shear mode and bulk mode --- two quantities that are
usually assumed to be uncorrelated --- at short times.

At longer times, $A$ and $B$ show different behavior. Graphene has a
characteristic density, which translates directly into a preferred
value for $A$ around which it fluctuates. The amplitude of the
fluctuations in $A$ are determined by the bulk modulus, which is an
equilibrium property and therefore computationally obtainable from
simulations without realistic dynamics. The aspect ratio $B$ does not
have an energetically preferred value, and its diffusive behavior is
therefore unrestricted. A practical consequence is that in simulations
the quantity $D_B$ can be determined more accurately than $D_A$, as
the latter shows a crossover from short-time diffusive behavior to
late-time saturation.

In our simulations, we have studied samples of polycrystalline
graphene with a variation in the amount of structural relaxation, the
size of the crystalline domains, and the density of structural defects
(mainly fivefold and sevenfold rings). In our simulations, we show a
linear relation between the number of such structural defects and the
diffusion coefficient $D_B$. In well-relaxed samples, large
crystalline domains are separated from each other by rows of
structural defects. Consequently, the number of defects decreases
linearly with the average domain size. We therefore expect also that
the diffusion coefficient $D_B$ decreases linearly with the average
domain size. In this context it will be useful to deepen this
connection to domain size engineering
\cite{li2010graphene,lin2019synthesis,zeng2020dynamically},
fabrication of polycrystalline graphene
\cite{milaninia2009all,paul2011production,mortazavi2014atomistic},
mechanics of grain boundaries
\cite{grantab2010anomalous,xu2016influence}.

From a materials science point of view, as well as from an experimental
point of view, the mechanical behavior of a sample of graphene under
external forces is important. We show that the deformation of graphene
under an external shear force is related to the quantity $D_B$ which is
readily accessible in simulations, via the Nerst-Einstein relation. For
this purpose, the external shear force is translated into a force $f_B$
on the quantity $B$, after which the shear rate $v_B=\partial B/\partial
t$ can be obtained from equation (\ref{eq7}), in which the diffusion coefficient
$D_B$ is used. And the mechanical deformation can then be readily obtained
from $v_B$.

We have limited ourselves to a relatively modest dynamical range of
$L_x$ and $L_y$, as well as relatively mild deformation
forces. Consequentially, in our simulations the domains do not get
deformed to elongated shapes but retain circular symmetry. If the
material would be stretched significantly in a time that is short
enough to rule out complete structural rearrangement, elongated
domains should arise, and the sample would experience restoring forces
back towards its original shape. This is illustrated in
Fig. \ref{fig6}(a). We speculate that this mechanism would actually
slow down the shearing process, making the shear distance non-linear
in time. Our simulations show signs of the onset of decreasing shear
rate in time [Fig. \ref{fig6}(d)]. A quantitative study of this
phenomenon, in which the possible relation between elongation of
domains and non-linear shear is investigated both in experiments and
mechanism, such as strengthening or weakening of graphene
\cite{Song2013,Yi2013,Huang2011}, fracture toughness
\cite{Han2017,dewapriya2018tailoring,jang2017uniaxial}, mechanical
mutability \cite{liu2014mechanical}, requires very long simulations,
which we will pick up in future work.  We believe these investigations
enhance our understanding of the mechanical properties of
polycrystalline graphene.

\begin{acknowledgments}
Z.L. acknowledges financial support from the China Scholarship Council (CSC)
\end{acknowledgments}

\end{document}